\documentclass[%
 reprint,
 amsmath,amssymb,
 aps,
]{revtex4-2}

\usepackage{xcolor}
\usepackage{graphicx}
\graphicspath{{./Figs/}}
\usepackage{dcolumn}
\usepackage{bm}

\begin{document}
\preprint{APS/123-QED}

\title{Dripping-onto-droplet capillary breakup}

\author{Ricardo El Khoury}
\thanks{ricardo.elkhoury@minesparis.psl.eu}
\affiliation{Mines Paris, PSL University, Centre for material forming (CEMEF), UMR CNRS 7635, rue  Claude Daunesse, 06904 Sophia-Antipolis, France}

\author{Kindness Isukwem}
\thanks{kindness-chinwendu.isukwem@minesparis.psl.eu}
\affiliation{Mines Paris, PSL University, Centre for material forming (CEMEF), UMR CNRS 7635, rue  Claude Daunesse, 06904 Sophia-Antipolis, France}

\author{Elie Hachem}
\thanks{elie.hachem@minesparis.psl.eu}
\affiliation{Mines Paris, PSL University, Centre for material forming (CEMEF), UMR CNRS 7635, rue  Claude Daunesse, 06904 Sophia-Antipolis, France}

\author{Anselmo Pereira}
\thanks{anselmo.soeiro\_pereira@minesparis.psl.eu (corresponding author)}
\affiliation{Mines Paris, PSL University, Centre for material forming (CEMEF), UMR CNRS 7635, rue  Claude Daunesse, 06904 Sophia-Antipolis, France}%

\date{\today}

\begin{abstract}
This experimental, numerical, and theoretical study investigates the capillary thinning and breakup of Newtonian filaments formed following the coalescence of a millimetric-nozzle-generated pendant drop with a lower droplet cap contained in a millimetric cylinder in ambient air, i.e., \textit{dripping-onto-droplet capillary breakup} (DoD). Our mixed approach combines filament breakup experiments recorded with a high-speed camera and three-dimensional numerical simulations based on a variational multiscale framework for multiphase fluid flows. The results are analysed by considering the dynamics of fluid filament thinning, energy transfers, and scaling laws. Three flow regimes are highlighted: capillary-inertial, capillary-viscous, and mixed capillary-inertial-viscous. All regimes are affected by gravity. The findings are summarised in a two-dimensional diagram that correlates the filament breakup time with different flow regimes using the important dimensionless parameters of the problem, e.g., the Ohnesorge number (which relates the viscous stress to inertial and capillary stresses) and the Bond number (which balances the gravitational stress with the capillary one). This diagram can be used to quantify both the liquid viscosity and the liquid-gas surface tension (for Newtonian fluids). Lastly, we demonstrate that DoD can also be used as a rheometric test, giving access to the extensional relaxation time of polymer solutions (for viscoelastic fluids).     
\vspace*{0.5cm}

\textbf{Keywords}: capillary thinning; extensional rheology; multiphase flow; experiments; 3D numerical simulation; scaling laws.
 
\end{abstract}

\maketitle

\section{Introduction} \label{INTRO}

The stretching and subsequent breakup of liquids are crucial in fluid mechanics. They are intrinsically related to critical environmental, everyday life, and industrial multiscale situations, such as the formation and fragmentation of large-scale cosmic sheets and filaments in galaxy clusters \citep{Kuchner_2021}, the optimisation of pesticide deposition via liquid atomisation \citep{Damak_2016}, the optimisation of liquid-propulsion systems \citep{Eggers_2008}, the airborne transmission of respiratory viruses through droplet emission during speech \citep{Abkarian_2020}, coating \citep{Weinstein_2004}, microfluidics \citep{Dewandre_2020}, inkjet printing \citep{He_2017}, 3D bioprinting \citep{Dey_2020, Yu_2025}, and, most importantly for the present study, extensional rheometry \citep{Liang_1994, Entov_1997, McKinley_2000, Tuladhar_2008, Keshavarz_2015, Zinelis_2024}. 

A vast number of extensional rheometry techniques rely on capillary stresses to favour the growth of Plateau-Rayleigh instabilities \citep{Plateau_1873, Rayleigh_1878, Rayleigh_1880, Rayleigh_1892}, leading to thinning, destabilisation, and breakup of the fluid filament. When the capillary pressure is counter-balanced by inertial or viscous stresses in a Newtonian fluid of density $\rho$, viscosity $\eta$ and surface tension $\sigma$, for instance, the material's diameter $d(t)$ theoretically decays with time $t$ as        
\begin{equation}
\frac{d(t)}{d_0} = A \left( \frac{t_b - t}{t_{c}} \right)^m  \, ,
\label{eq:intro-1}
\end{equation} 
in which $d_0$ is the initial filament diameter, $t_b$ is the instant at which the filament breaks apart, and $A$, $m$ and $t_c$ are, respectively, a prefactor, an exponent and a characteristic time related to stresses dominating the thinning process: $A \approx 1.434$, $m = 2/3$ and $t_c = \sqrt{\rho d_0^3/\sigma}$ in the capillary-inertial regime \citep[emerging from a balance between capillary and inertial stresses in inviscid fluids][]{Peregrine_1990, Eggers_1993}; and $A \approx 0.1418$, $m = 1$ and $t_c = \eta d_0/\sigma$ in the capillary-viscous regime \citep[resulting from a competition between capillary and viscous stresses in the gravity-free and inertialess limit;][]{Papageorgiou_1995, McKinley_2000}. Furthermore, dilute solutions of extensible polymers (with $\lambda_e/t_c > 0$, where $\lambda_e$ is the solution's extensional relaxation time) exhibit a capillary-elastic thinning dynamics (given by a competition between capillary and elastic stresses in absence of gravity and inertia; e.g., capillary-elastic regime) with a filament diameter that decays exponentially in time according to the following equation derived by Entov and Hinch \citep{Entov_1997} for an Oldroyd-B/Hookean dumbbell \citep{Oldroyd_1950, Bird_1987, Hinch_2021}:           
\begin{equation}
\frac{d(t)}{d_0} = A \left[\frac{\left(\eta_0 - \eta_s \right) d_0}{\sigma \lambda_e} \right]^{1/3} \exp{\left( -\frac{t}{3\lambda_e} \right)}  \, ,
\label{eq:intro-2}
\end{equation} 
where $\eta_0$ is the solution's zero-shear viscosity, $\eta_s$ is the solvent viscosity, $\eta_0 - \eta_s$ is the polymeric viscosity $\eta_p$, and $A=(1/4)^{1/3}$ \citep{Bazilevskii_1997, Clasen_2006a, Pereira_2013}. Hence, one could use equations \ref{eq:intro-1} and \ref{eq:intro-2} to access material properties such as viscosity (for Newtonian fluids), surface tension, and relaxation time (for viscoelastic fluids). 
\begin{figure*}[htp]
\centering    
\includegraphics[width=1\linewidth]{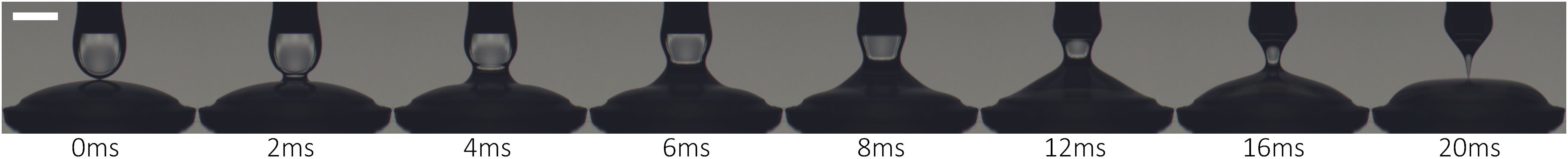}
\vspace{-0.7cm}    
\caption{{\color{darkgray}{Snapshots illustrating the capillary-thinning process in a 0.01Pa$\cdot$s-viscosity silicone filament generated after the coalescence of a pendant droplet attached to a nozzle and a lower droplet cap contained in a millimetric cylinder in ambient air under gravity (liquid density $\rho = 1000$kg/m$^3$, and surface tension $\sigma$ = 0.02N/m). White scale bar = 1.37mm. The filament breaks apart in 20ms ($t_b = 20$ms).}}} 
\vspace{-0.3cm} 
\label{fig-1}
\end{figure*}

The above theoretical predictions have been non-exhaustively compared with experimental results obtained from a variety of flow configurations, including gravity-induced drop formation \citep[e.g., dripping from a nozzle;][]{Eggers_1993, Eggers_1997, Eggers_2005, Javadi_2013, Deblais_2018, Dinic_2019, Bazazi_2025}, liquid bridges \citep[Capillary Breakup Extensional Rheometer (CaBER); Dripping-onto-Substrate (DoS); Acoustically-Driven Microfluidic Extensional Rheometer (ADMiER);][]{Gaudet_1996, Anna_2001, Rodd_2005, Clasen_2006b, Dinic_2015, Dinic_2017, Rosello_2019, Valette_2019, Joseph_2025, Bhattacharjee_2011, McDonnell_2015}, and jetting \citep[liquid jets; Rayleigh Ohnesorge Jetting Extensional Rheometer (ROJER)][]{Bogy_1979, Schummer_1983, Eggers_2008, Keshavarz_2015, Greiciunas_2017, Mathues_2018, Kooij_2025}. The use of such flows is, nevertheless, frequently constrained by numerous issues arising from their technical characteristics, leading to discrepancies between theoretical predictions and experimental findings. Dripping, for example, tends to be affected both by viscous stresses \citep[even for low viscosity fluids;][]{Deblais_2018} and secondary inertial stresses resulting from the acceleration of the falling drop during the gravity-induced stretching \citep{Dinic_2019}. The generation of liquid bridges, in turn, often involves an initial step-strain promoted by wetting and axial displacement of moving plates (or pistons), which can also lead to contamination by inertial stresses, as well as deformation history and contact angle effects \citep{Zinelis_2024, Hu_2025}. Furthermore, capillary thinning in liquid bridges is affected by gravity even in millimetric filaments, which modifies the value of the prefactor $A$ \citep{Zinelis_2024} and consequently makes the use of equations \ref{eq:intro-1} and \ref{eq:intro-2} for rheological purposes difficult. Finally, since liquid jets rely on the growth of instabilities throughout an axially-perturbed filament \citep{Keshavarz_2015, Keshavarz_2016, Greiciunas_2017}, they are highly affected by axial tension \citep{Mathues_2018}, which leads to a diameter's exponential decay different from that predicted by equation \ref{eq:intro-2}.                

To overcome the technical constraints associated with the prototypical configurations mentioned above, we focus on a flow conformation called here \textit{dripping onto droplet} (DoD) and illustrated in figure \ref{fig-1}: the capillary-thinning and subsequent breakup of a fluid filament formed due to the coalescence of an upper millimetric pendant drop generated through a nozzle with a lower droplet cap of the same fluid contained in a millimetric cylinder in ambient air. In such a wetting-independent scenario, the capillary thinning process naturally occurs when the upper drop gently touches the lower one (no fast pre-stretching required). The absence of axial displacement mitigates secondary inertial stresses. At the same time, gravity effects are dramatically attenuated by reducing the diameter of the upper droplet (no gravity needed to initiate the flow), which ultimately favours the development of a thinning process more in line with the scenarios described by equations \ref{eq:intro-1} (for Newtonian fluids) and \ref{eq:intro-2} (for viscoelastic fluids). In summary, from a practical standpoint, the DoD setup requires only standard components: a syringe pump and a millimetric nozzle (to generate the upper pendant droplet), a 3D-printed cylindrical pool (to restrict the lateral movement of the lower droplet cap), a light source, and a high-speed camera (see the schematic in figure \ref{fig-2}\textit{a}). Such practicality makes the DoD a simple-to-assemble, easy-to-use rheometric test.

Aiming to shed light on the possibilities and new avenues DoD offers as an extension-based rheometric tool, we present a mixed experimental, numerical and theoretical study focused on the physical mechanisms governing the stretching and the subsequent breakup process in a Newtonian fluid filament generated after the coalescence of an upper millimetric droplet and a lower droplet cap. Our mixed approach combines filament breakup experiments recorded by a high-speed camera with three-dimensional (3D) numerical simulations based on a variational multiscale framework for multiphase fluid flows. The results are depicted through stretching dynamics analyses, energy transfers, and scaling laws. Additionally, to extend the classical approach based on the time-evolution of the filament diameter, we focus on the connections between the breakup time and the physical mechanisms driving the thinning process. The findings are summarised in a two-dimensional diagram that correlates $t_b$ with different flow regimes using the important dimensionless parameters of the problem, e.g., the Ohnesorge number (which relates the viscous stress to inertial and capillary stresses) and the Bond number (which balances the gravitational stress with the capillary stress). This diagram allows us to quantify both the liquid viscosity and the liquid-air surface tension. Lastly, we briefly demonstrate that DoD can also be extended to non-Newtonian fluids and give access to the extensional relaxation time of polymer solutions.        

The organisation of the paper is as follows. A detailed description of the physical formulation and the mixed experimental-numerical method is presented in section \ref{PFENMDN}. The important parameters of the problem are equally highlighted. Experimental and numerical results are discussed in section \ref{RD}. Finally, conclusions and perspectives are drawn in the closing section.

\section{Physical Formulation, Experimental-Numerical Method, and Dimensionless Numbers} \label{PFENMDN}

\begin{figure*}[htp]
\centering    
\includegraphics[width=1\linewidth]{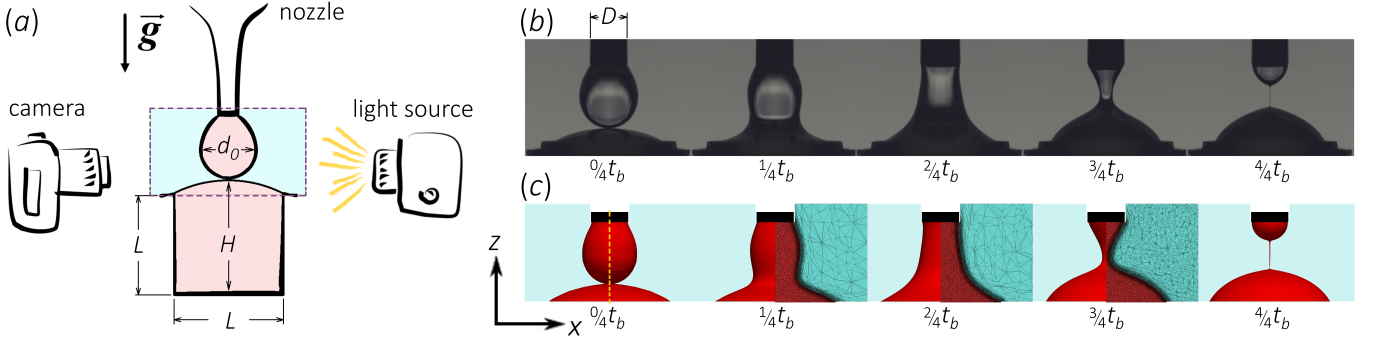}
\vspace{-0.7cm}    
\caption{{\color{darkgray}{(\textit{a}) Schematic illustration of the experimental set-up for the investigation of the capillary thinning and breakup of Newtonian filaments formed following the coalescence of a millimetric-nozzle-generated pendant drop (of density $\rho$, viscosity $\eta$, surface tension $\sigma$ and diameter $d_0$) with a lower droplet cap contained in a 3D-printed-polylactic-acid cylindrical pool of millimetric diameter $L$ and height $L$ in ambient air under gravity, i.e., \textit{dripping-onto-droplet capillary breakup} (DoD). The blue rectangle, delimited by the magenta-dashed lines, highlights the free-surface flow region. (\textit{b}) Typical snapshots captured by the high-speed camera illustrating the capillary-thinning process from the droplets' coalescence until the filament breakup at instant $t_b$. (\textit{c}) Snapshots showing the time evolution of a typical numerical simulation from the droplets' coalescence until the filament breakup at instant $t_b$. The used mesh (composed of approximately $10^6$ elements) is adapted around each interface (liquid in red and surrounding air in blue), as illustrated by the black lines on the right side from the second to the fourth snapshot. Note that only the free-surface flow part is displayed in (\textit{b}) and (\textit{c}). Lastly, cylindrical pool dimensions $L \times L$, the distance $H$ between the bottom of the pool and the top of the lower droplet, and the nozzle diameter $D$ are kept fixed for all the experiments and numerical simulations performed here ($L = 4.35$mm, $H = 5.1$mm, $D = 1.37$mm).}}}
\vspace{-0.2cm} 
\label{fig-2}
\end{figure*}

As mentioned in the previous section, this study presents a combined experimental, numerical and theoretical analysis of the capillary-driven filament thinning process followed by the coalescence between an upper Newtonian droplet of density $\rho$, viscosity $\eta$, and diameter $d_0$ and a lower droplet of the same fluid, as depicted in figure \ref{fig-2}. The former is generated by a nozzle of millimetric diameter $D$, while the latter is formed when filling a 3D-printed-polylactic-acid cylindrical pool of millimetric diameter and height $L \times L$ (see figure \ref{fig-2}\textit{a}). The droplets' coalescence leads to the formation of a fluid filament whose diameter $d(t)$ progressively decreases under capillary forces until its rupture at instant $t_b$ due to the growth of Plateau-Rayleigh instabilities \citep{Eggers_1993, Eggers_2008, Pita_2012, Deblais_2018, Kooij_2025}. The surrounding air is characterised by a density $\rho_{air}$ and viscosity $\eta_{air}$. Both liquid and gas phases are at fixed ambient temperature $T$, and the surface tension between them $\sigma$ is constant. 

As schematised in figure \ref{fig-2}(\textit{a}), the capillary-thinning process is recorded by an Optronis Cyclone 2-2000 high-speed camera operated at $\mathcal{O} \left( 10^{4} \right)$ frames/s and equipped with a Sigma 105mm F2.8 DG OS HSM macro lens. A LED backlight system provides the necessary lighting level. A typical experimental snapshot sequence illustrating the fluid filament breakup is given in figure \ref{fig-2}(\textit{b}). Deionised water ($\rho = 999$kg/m$^3$, $\eta = 10^{-3}$Pa.s, $\sigma = 72$mN/m), glycerol-water mixtures ($1050$kg/m$^3$ $\leq \rho \leq 1260$kg/m$^3$, $0.00105$Pa.s $\leq \eta \leq  1.91$Pa.s, 68mN/m $\leq \sigma \leq$ 71mN/m), and silicone oils ($913$kg/m$^3$ $\leq \rho \leq 970$kg/m$^3$, $0.0052$Pa.s $\leq \eta \leq  24.5$Pa.s, $\sigma = 20$mN/m) are used. Initial upper droplet diameters $d_0$ are in a range of 1.5mm-2.6mm. Additionally, the nozzle diameter $D$ ranges from 1.37mm to 1.87mm, and pool's dimensions (diameter and height) $L$ are kept fixed at 4.35mm. Lastly, the distance $H$ between the bottom of the pool and the top of the lower droplet is kept at 5.1mm for all flow cases explored here. A movie showing a standard experiment is available on {\color{purple}{https://www.youtube.com/watch?v=ctmVaL1ZbV8}}.

Our computational method is based on a massively parallel finite element library devoted to multiphase flows \citep[CIMLIB-CFD;][]{Valette_2019, Pereira_2019, Pereira_2020, Valette_2021, Isukwem_2024a, Isukwem_2024b, Isukwem_2024c, Isukwem_2025a, Isukwem_2025b, Isukwem_2025c, Isukwem_2025d, Godefroid_2025}. More specifically, we apply the momentum conservation equation presented below to the solenoidal flow ($\boldsymbol{\nabla \cdot u} = 0$) described earlier (see Fig. \ref{fig-1}):     
\begin{equation}
\rho \left( \frac{\partial \boldsymbol{u}}{\partial t} + \boldsymbol{u} \cdot \nabla \boldsymbol{u} - \boldsymbol{g} \right) = - \nabla p + \nabla \cdot \boldsymbol{\tau} + \boldsymbol{f_{st}} \, ,
\label{eq:cons-mom}
\end{equation} 
in which $\boldsymbol{u}$, $p$, $\boldsymbol{\tau}$, $\nabla$, $\boldsymbol{g}$, $\nabla \cdot$ and $\boldsymbol{f_{st}}$ are, respectively, the velocity vector $\boldsymbol{u} = \left\lbrace u_x, u_y, u_z \right\rbrace $, the pressure, the extra-stress tensor, the gradient operator, the gravity vector, the divergence operator, and a capillary term related to the surface tension force. The latter is defined as $\boldsymbol{f_{st}} = -\sigma \kappa \Phi \boldsymbol{n}$, where $\sigma$, $\kappa$, $\Phi$, and $\boldsymbol{n}$ are the sheet surface tension with the surrounding air, the curvature of the sheet-air interface, the Dirac function locating the sheet-air interface, and its normal vector, respectively. In addition, the extra-stress tensor is given by $\boldsymbol{\tau} = \eta \boldsymbol{\dot{\gamma}}$, in which $\boldsymbol{\dot{\gamma}}$ represents the \textit{rate-of-strain} tensor defined as $\boldsymbol{\dot{\gamma}} = \left( \boldsymbol{\nabla u} + \boldsymbol{\nabla u}^T \right)$. The norm of $\boldsymbol{\dot{\gamma}}$ is called \textit{deformation rate}, being defined as $|\boldsymbol{\dot{\gamma}}| = \left( \frac{1}{2} \boldsymbol{\dot{\gamma}}: \boldsymbol{\dot{\gamma}} \right)^{\frac{1}{2}}$. A typical numerical snapshot sequence showing the fluid filament breakup is shown in figure \ref{fig-2}(\textit{c}), where the liquid phase is represented in red and the surrounding gas phase in blue. 

The numerical methods employed involve a \textit{Variational Multiscale Method} (\textit{VMS}) coupled with an anisotropic mesh adaptation technique \citep{Valette_2019, Pereira_2019, Pereira_2020}. The computational meshes consist of approximately 10$^{6}$ elements, with a minimum size of 1$\mu$m and a maximum size of $d_0/10$, as illustrated by the black lines in figure \ref{fig-2}(\textit{c}). A computation box of dimension $5L$ $\times$ $5L$ $\times$ $5L$ is used (not shown for brevity). The Courant--Friedrichs--Lewy (CFL) condition is lower than 0.1 for all numerical simulations considered here. The evolution of the liquid interface over time is tracked using a Level-Set function \citep{Isukwem_2025d}. Fluid properties are interpolated through the Level-Set function whose value varies from positive within the sheet to negative in the air \citep[and zero at the interfaces between the different fluid phases; supplemental details concerning the interface capturing and the Level-Set function are available in][]{Isukwem_2025a, Isukwem_2025d}.  

Concerning the boundary and initial conditions, a fully developed Poiseuille (parabolic) profile is imposed as the liquid downwards vertical velocity at the nozzle inlet 
\begin{equation}
u_{z,inlet} = 2U_0\left[ 1 - \left(  \frac{r}{R}  \right)^2  \right]   \, ,
\label{eq:U0}
\end{equation} 
where $U_0$ denotes the liquid mean vertical velocity at the nozzle inlet, $r$ is the horizontal distance from the centerline (yellow-dashed line in figure \ref{fig-2}\textit{c}; $r^2 = x^2 + y^2$), and $R$ is the nozzle's radius ($R = D/2$). Furthermore, a no-slip condition is imposed between the liquid and the walls of the nozzle, as well as between the liquid and the cylindrical pool. Zero normal stresses are imposed on the other walls of the computational box. Lastly, each liquid's initial free-surface shape is directly extracted from the corresponding experimental case at the droplets' coalescence instant (for example, the liquid's experimental morphology in the first snapshot in figure \ref{fig-2}\textit{b} is merged into the numerical code and used as the liquid's initial shape at $t = 0$s). 

It is important to emphasize that our numerical multiphase framework has been validated for a variety of flow scenarios, which includes the impact of drops \citep{Isukwem_2024a, Isukwem_2024b, Isukwem_2024c, Isukwem_2025a, Isukwem_2025c, Godefroid_2025}, dam breaks \citep{Valette_2021, Isukwem_2025b}, the stretching and breakup of filaments and sheets \citep{Valette_2019, Isukwem_2025d}, and the development of buckling instabilities \citep{Pereira_2019, Pereira_2020} using Newtonian and non-Newtonian fluids. Supplemental DoD-based experimental-numerical comparisons are also provided in the following section. 

The dimensionless numbers governing the problem, derived from the Buckingham-$\Pi$ theorem, are based on the dimensional variables $d_0$, $\rho$, $\eta$, $\sigma$, $g$ (the $z$ component of the gravity vector), $U_0$, $\rho_{air}$ and $\eta_{air}$, with the fundamental units of mass [kg], distance [m] and time [s]. These variables lead to five important dimensionless quantities:
\begin{equation}
\Pi_1 = \frac{3\eta \sqrt{\sigma/(\rho d_0^3)}}{\sigma/d_0}  \, ,
\label{eq2.1}
\end{equation}
\begin{equation}
\Pi_2 = \frac{\rho g d_0}{\sigma/d_0}  \, ,
\label{eq2.2}
\end{equation}
\begin{equation}
\Pi_3 = \frac{\rho U_0^2}{\sigma/d_0}  \, ,
\label{eq2.3}
\end{equation}
\begin{equation}
\Pi_4 = \frac{\eta_{air} \sqrt{\sigma/(\rho d_0^3)}}{\sigma/d_0}   \, ,
\label{eq2.4}
\end{equation}
\begin{equation}
\Pi_5 =  \frac{\rho_{air}}{\rho}   \, .
\label{eq2.5}
\end{equation}
$\Pi_1$ represents the Ohnesorge number ($\mathrm{Oh}$; the prefactor 3 translates the uniaxial extensional nature of the flows, a point deeper discussed in the following section), $\Pi_2$ corresponds to the Bond number ($\mathrm{Bo}$), $\Pi_3$ is the Weber number based on the liquid's mean velocity into the slot ($\mathrm{We_{slot}}$), $\Pi_4$ is the air-based Ohnesorge number ($\mathrm{Oh_{air}}$), $\Pi_5$ is the density ratio, and $\sqrt{\sigma/(\rho d_0^3)}$ is a characteristic strain rate associated with the capillary-thinning process. In the present work, ${\mathrm{Oh_{air}}} \lesssim 10^{-5}$, $\rho_{air}/\rho \lesssim 10^{-3}$, and $\mathrm{We}_{slot} \lesssim 10^{-10}$ ($U_0$ is much lower than the characteristic stretching velocity $\sqrt{\sigma/(\rho d_0^3)}$). Thus, the flow scenarios considered here vary with only two key dimensionless numbers:  
\begin{equation}
\mathrm{Oh} = \frac{3\eta \sqrt{\sigma/(\rho d_0^3)}}{\sigma/d_0}  ~~~ \text{(Ohnesorge number)}\, .
\label{eq2.6}
\end{equation}
\begin{equation}
\mathrm{Bo} = \frac{\rho g d_0}{\sigma/d_0}  ~~~ \text{(Bond number)}\, .
\label{eq2.7}
\end{equation}
The Ohnesorge number ranges from 0.001 to 1000, while the Bond number varies from 0 to 3. The impact of these two dimensionless numbers on the stretching and rupture of Newtonian filaments is discussed in section \ref{RD}, where the results are presented in terms of these quantities. 

\section{Results and Discussion} \label{RD}
\begin{figure*}[htp]
\centering    
\includegraphics[width=1\linewidth]{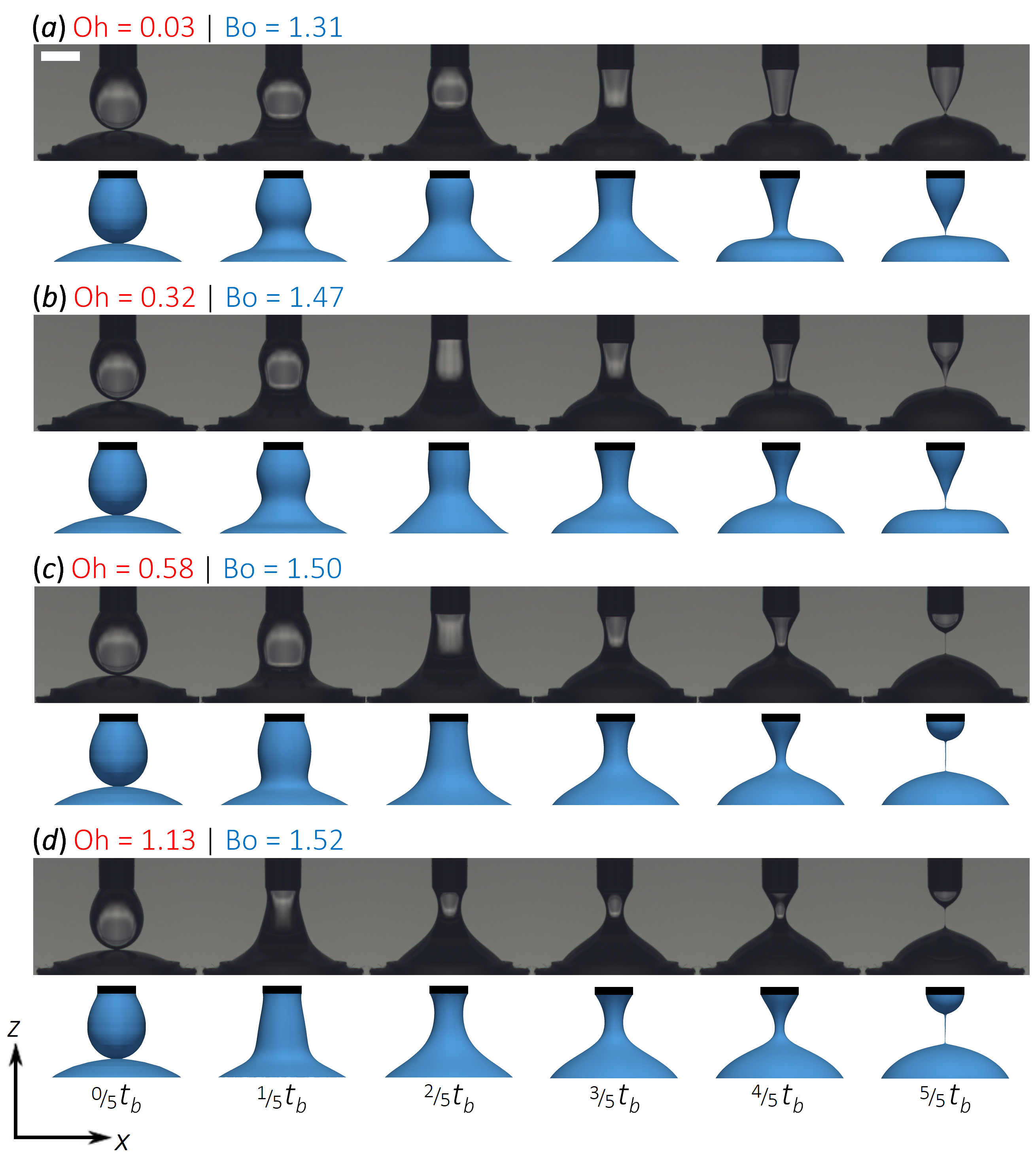}
\vspace{-0.7cm}    
\caption{{\color{darkgray}{Droplets' coalescence followed by the capillary thinning and breakup of the formed filaments at four $\mathrm{Oh}$-$\mathrm{Bo}$ couples: (\textit{a}) $\mathrm{Oh} = 0.03$ and $\mathrm{Bo} = 1.31$; (\textit{b}) $\mathrm{Oh} = 0.32$ and $\mathrm{Bo} = 1.47$; (\textit{c}) $\mathrm{Oh} = 0.58$ and $\mathrm{Bo} = 1.50$; and (\textit{d}) $\mathrm{Oh} = 1.13$ and $\mathrm{Bo} = 1.52$. The upper row of each sub-figure is composed of six successive glycerol-in-water snapshots showing the time evolution of the thinning process until the filaments' breakup, while the corresponding numerical images (in blue) are displayed in the lower row. Note that only the free-surface part of the liquid is depicted. The time interval between subsequent snapshots is $1/5t_b$. White scale bar = 1.37mm.}}} 
\vspace{-0.2cm} 
\label{fig-3}
\end{figure*}

Figure \ref{fig-3} illustrates droplets' coalescence followed by the capillary thinning and breakup of the formed filaments at four $\mathrm{Oh}$-$\mathrm{Bo}$ couples: $\mathrm{Oh} = 0.03$ and $\mathrm{Bo} = 1.31$ (figure \ref{fig-3}\textit{a}); $\mathrm{Oh} = 0.32$ and $\mathrm{Bo} = 1.47$ (figure \ref{fig-3}\textit{b}); $\mathrm{Oh} = 0.58$ and $\mathrm{Bo} = 1.50$ (figure \ref{fig-3}\textit{c}); $\mathrm{Oh} = 1.13$ and $\mathrm{Bo} = 1.52$ (figure \ref{fig-3}\textit{d}). Since the Bond number slightly varies from 1.31 to 1.52, figure \ref{fig-3} primarily stresses the Ohnesorge number effects on the capillary thinning processes. The top row of each sub-figure is composed of six successive glycerol-in-water snapshots showing the time evolution of the thinning process until the filaments' breakup, while the corresponding numerical images (in blue) are displayed in the bottom row. Note that only the free-surface part of the liquid is depicted. The time interval between subsequent snapshots is $1/5t_b$. 

\begin{figure*}[htp]
\centering    
\includegraphics[width=1\linewidth]{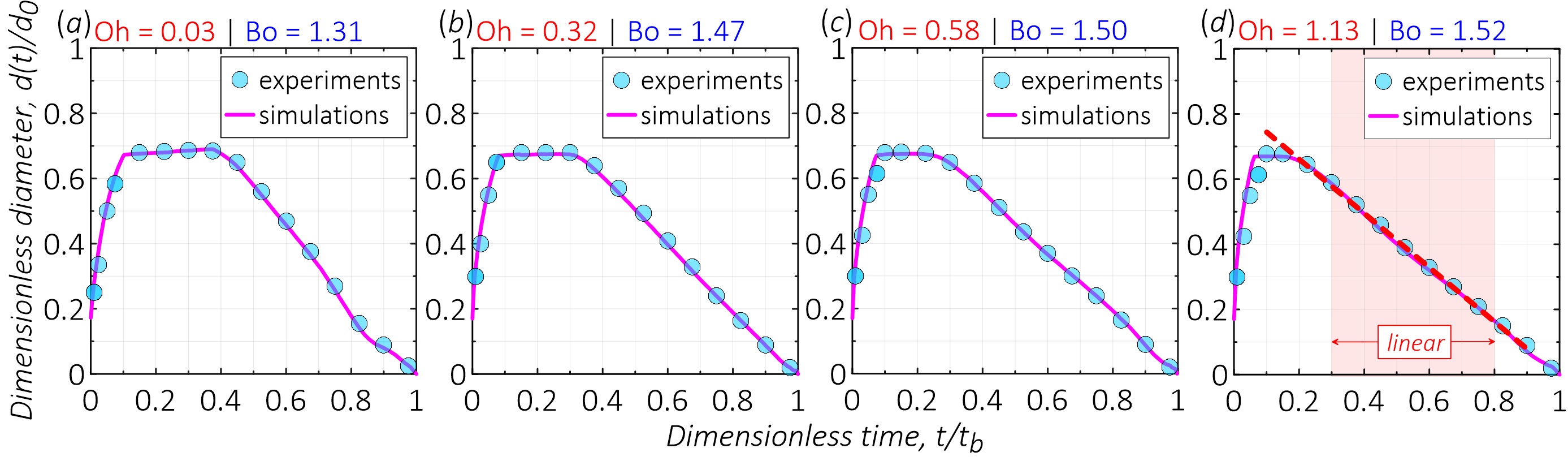}
\vspace{-0.7cm}    
\caption{{\color{darkgray}{Experimental and numerical results showing the instantaneous minimum dimensionless diameter $d(t)/d_0$ as a function of the dimensionless time $t/t_b$ of liquid filaments at four $\mathrm{Oh}$-$\mathrm{Bo}$ couples: (\textit{a}) $\mathrm{Oh} = 0.03$ and $\mathrm{Bo} = 1.31$; (\textit{b}) $\mathrm{Oh} = 0.32$ and $\mathrm{Bo} = 1.47$; (\textit{c}) $\mathrm{Oh} = 0.58$ and $\mathrm{Bo} = 1.50$; and (\textit{d}) $\mathrm{Oh} = 1.13$ and $\mathrm{Bo} = 1.52$. The sky-blue circles represent the experimental data, while the solid magenta lines denote the simulations.}}} 
\vspace{-0.2cm} 
\label{fig-4}
\end{figure*}

\begin{figure}[htp]
\centering    
\includegraphics[width=1\linewidth]{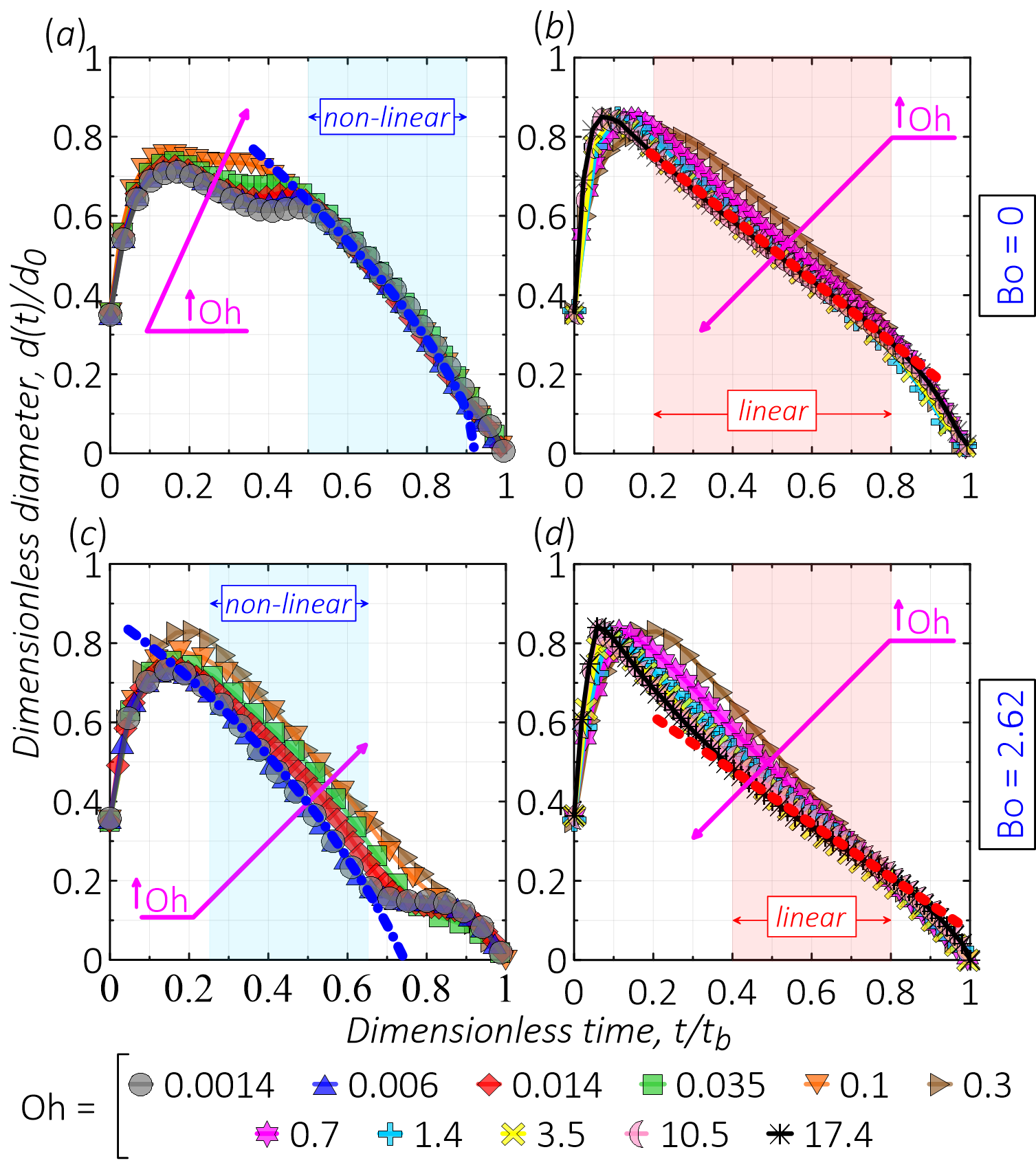}
\vspace{-0.7cm}    
\caption{{\color{darkgray}{Numerical results showing $d(t)/d_0$ against $t/t_b$ of liquid filaments at $\mathrm{Bo} = 0$ (top row) and $\mathrm{Bo} = 2.62$ (bottom row) for eleven different $\mathrm{Oh}$: 0.0014, 0.006, 0.014, 0.035, and 0.1 (left column sub-figures; \textit{a} and \textit{c}); and 0.3, 0.7, 1.4, 3.5, 10.5, and 17.4 (right column sub-figures; \textit{b} and \textit{d}). A specific solid symbol denotes each numerical flow case, and the increase of $\mathrm{Oh}$ in each sub-figure is highlighted by the magenta arrows. The blue-dotted-dash line in the blue region denotes a diameter decay proportional to $(1-t/t_b)^{2/3}$, while the red-dashed line in the red region represents a linear diameter proportional to $(1-t/t_b)$.}}} 
\vspace{-0.2cm} 
\label{fig-5}
\end{figure}

At $\mathrm{Oh} = 0.03$, the filament diameter evolves in a non-uniform way along the $z$ direction, leading to a liquid conical shape at $t_b$. Nevertheless, such heterogeneity tends to vanish as $\mathrm{Oh}$ increases, thereby amplifying the viscous stresses relative to the inertial and capillary stresses, leading to a homogeneous centre filament that uniformly thins until it breaks up. Logically, higher viscosities delay the growth of perturbations within the liquid, favouring the formation of a homogeneous liquid bridge between the upper and lower droplets. Uniform centre filaments are typically observed at $\mathrm{Oh} \ge 1$.                

The instantaneous minimum dimensionless diameter $d(t)/d_0$ of the filaments illustrated in figures \ref{fig-3}(\textit{a})-(\textit{d}) are respectively plotted against $t/t_b$ in figures \ref{fig-4}(\textit{a})-(\textit{d}). The sky-blue circles represent the experimental data, while the solid magenta lines denote the simulations. The initial diameter increase observed in the curves results from the coalescence of initial droplets and the formation of a liquid filament, followed by capillary thinning. In the latter stage, a non-linear diameter decay ($t/t_b > 0.45$) is clearly noted at $\mathrm{Oh} = 0.03$, which evolves towards a more linear profile as $\mathrm{Oh}$ becomes more pronounced (e.g., a red dashed straight line fits the data for $0.3 \leq t/t_b \leq 0.8$ at $\mathrm{Oh} = 1.13$).                 

The good agreement between experiments and numerical simulations, as shown in the previous figures, gives us confidence to numerically explore the effects of the Ohnesorge number on capillary thinning dynamics. In these connections, we plot in figure \ref{fig-5} $d(t)/d_0$ against $t/t_b$ at $\mathrm{Bo} = 0$ (sub-figures \ref{fig-5}\textit{a}-\textit{b}; gravity free) and $\mathrm{Bo} = 2.62$ (sub-figures \ref{fig-5}\textit{c}-\textit{d}; comparable gravitational and capillary stresses) for eleven different $\mathrm{Oh}$: 0.0014, 0.006, 0.014, 0.035, and 0.1 (left column sub-figures); and 0.3, 0.7, 1.4, 3.5, 10.5, and 17.4 (right column sub-figures). Each solid symbol denotes a specific numerical flow case, and the magenta arrows highlight the increase of $\mathrm{Oh}$ in each sub-figure. 

Under gravity-free conditions ($\mathrm{Bo} = 0$; sub-figures \ref{fig-5}\textit{a}-\textit{b}), we first observe that the diameter profiles exhibit an initial oscillatory stage ($t/t_b < 0.5$) followed by a non-linear decay ($t/t_b > 0.5$) at $\mathrm{Oh} \le 0.006$. Logically, the oscillation is mitigated by the strengthening of the viscous stresses relative to the inertial and the capillary stresses, and thus the augmentation of $\mathrm{Oh}$. Moreover, for $\mathrm{Oh} \le 0.1$, the $d(t)/d_0$ profiles exhibit a similar non-linear decay, as underlined by the blue region and the blue dash-dotted lines. A linear diameter decay emerges as $\mathrm{Oh}$ becomes more pronounced, which is in line with sub-figure \ref{fig-4}(\textit{d}) and associated with the uniform thinning process depicted in sub-figure \ref{fig-3}(\textit{d}). The red dashed straight lines highlight this linear decay. 

Contrasting profiles emerge under comparable gravitational and capillary stresses (sub-figures \ref{fig-5}\textit{c}-\textit{d}), which reveal non-marginal gravity effects on the capillary thinning dynamics even at a relatively low Bond number. Differences between the $\mathrm{Bo} = 0$ and $\mathrm{Bo} = 2.62$ curves are observed for the whole range of Ohnesorge number considered in figure \ref{fig-5}. At $\mathrm{Oh} \le 0.01$ for instance, the increase of $\mathrm{Bo}$ anticipates the non-linear filament diameter decay (note that the beginning of the diameter decay moves from $t/t_b = 0.5$ in sub-figure \ref{fig-5}\textit{a} to $t/t_b = 0.25$ in sub-figure \ref{fig-5}\textit{c}). Additionally, at $\mathrm{Oh} \ge 10$, the increase of $\mathrm{Bo}$ shortens the linear part of the curves, as highlighted by the red regions in the graphs. 

Finally, reading figure \ref{fig-5}, it is essential to emphasise that, for a fixed $\mathrm{Bo}$, the same $d(t)/d_0$ profiles are obtained for $\mathrm{Oh} \le 0.01$ and $\mathrm{Oh} \ge 10$ (note the grey circle and the blue triangle curves in sub-figures \ref{fig-5}\textit{a} or \ref{fig-5}\textit{c}, as well as the pink croissant and the black asterisk curves in sub-figures \ref{fig-5}\textit{b} or \ref{fig-5}\textit{d}). Such finds suggest the existence of borderline regimes for $\mathrm{Oh} << 1$ and $\mathrm{Oh} >> 1$, at least for the range of Bond number explored here ($0 \le \mathrm{Bo} < 3$).               

\begin{figure*}[htp]
\centering    
\includegraphics[width=1\linewidth]{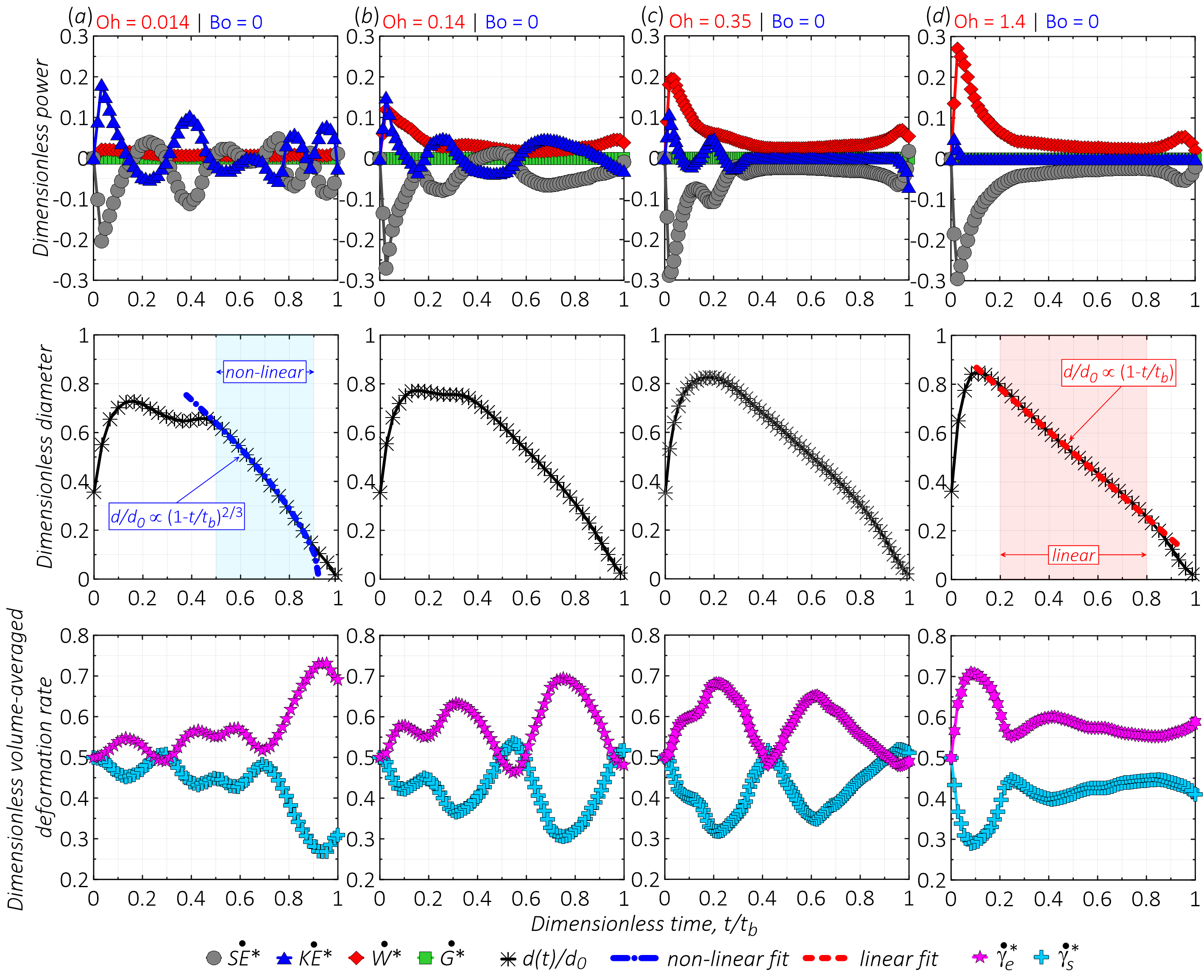}
\vspace{-0.7cm}    
\caption{{\color{darkgray}{Numerical results showing the dimensionless energy transfer (top row), the minimum dimensionless diameter (black asterisks in the middle row), and the dimensionless volume-averaged deformation rate (bottom row) associated with extension/compression (${\dot{\gamma}}^{\ast}_{e}$; pink stars), and shear (${\dot{\gamma}}^{\ast}_{s}$; cyan crosses) during the filament thinning at $\mathrm{Bo} = 0$ (gravity free) for four $\mathrm{Oh}$: (\textit{a}) 0.014; (\textit{b}) 0.14; (\textit{c}) 0.35; and (\textit{d}) 1.4. The blue-dotted-dash line in the flow region denotes a diameter decay proportional to $(1-t/t_b)^{2/3}$, while the red-dashed line in the red region represents a linear diameter decay proportional to $(1-t/t_b)$.}}}
\vspace{-0.2cm}  
\label{fig-6}
\end{figure*}

\begin{figure*}[htp]
\centering    
\includegraphics[width=1\linewidth]{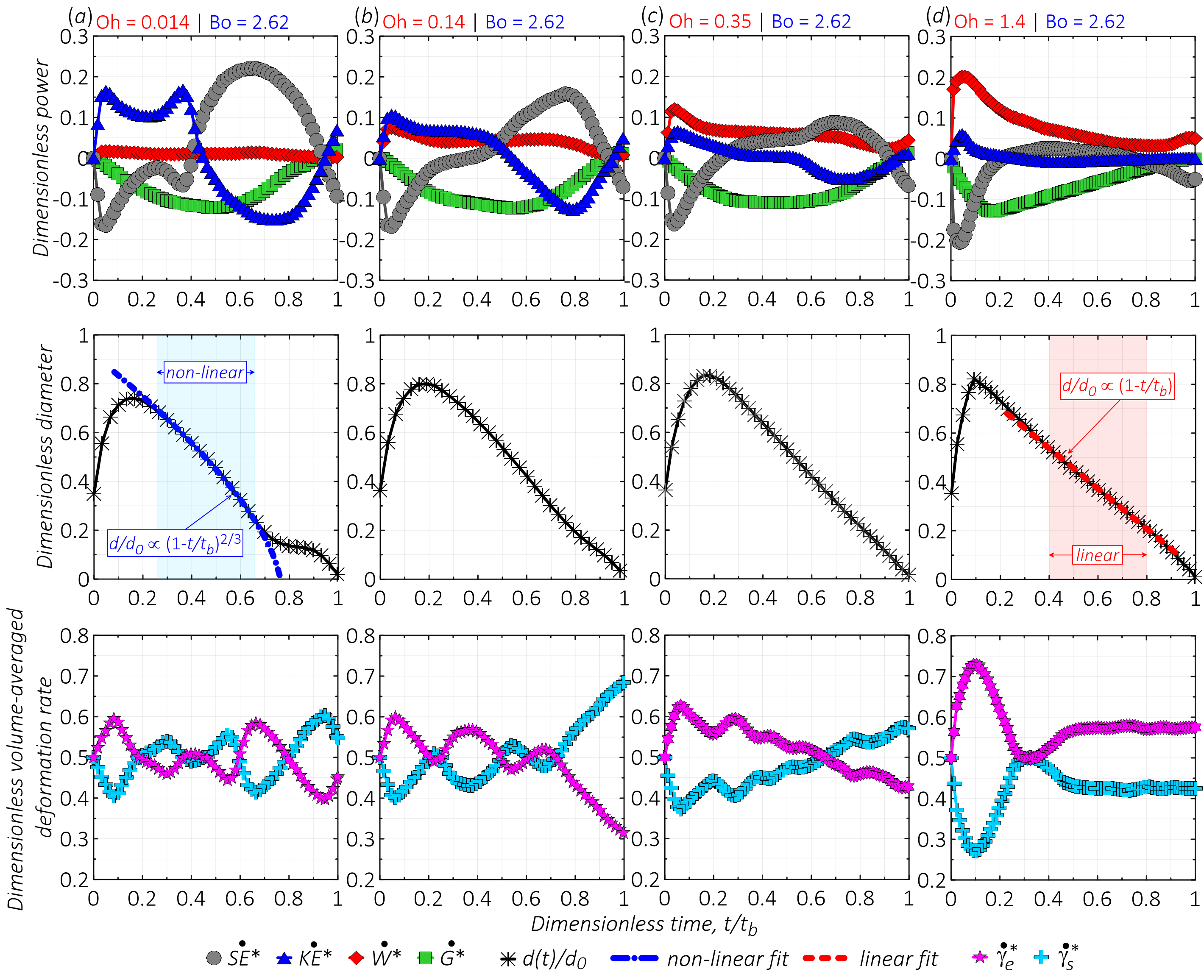}
\vspace{-0.7cm}    
\caption{{\color{darkgray}{Numerical results showing the dimensionless energy transfer (top row), the minimum dimensionless diameter (black asterisks in the middle row), and the dimensionless volume-averaged deformation rate (bottom row) associated with extension/compression (${\dot{\gamma}}^{\ast}_{e}$; pink stars), and shear (${\dot{\gamma}}^{\ast}_{s}$; cyan crosses) during the filament thinning at $\mathrm{Bo} = 2.62$ for four $\mathrm{Oh}$: (\textit{a}) 0.014; (\textit{b}) 0.14; (\textit{c}) 0.35; and (\textit{d}) 1.4. The blue-dotted-dash line in the flow region denotes a diameter decay proportional to $(1-t/t_b)^{2/3}$, while the red-dashed line in the red region represents a linear diameter decay proportional to $(1-t/t_b)$.}}}
\vspace{-0.2cm}  
\label{fig-7}
\end{figure*}

To gain a deeper understanding of the physical mechanisms driving the thinning process, figures \ref{fig-6} and \ref{fig-7} present the dimensionless energy transfer during the filament thinning respectively at $\mathrm{Bo} = 0$ and $\mathrm{Bo} = 2.62$ for four $\mathrm{Oh}$: 0.014 (sub-figures \ref{fig-6}\textit{a} and \ref{fig-7}\textit{a}), 0.14 (sub-figures \ref{fig-6}\textit{b} and \ref{fig-7}\textit{b}), 0.35 (sub-figures \ref{fig-6}\textit{c} and \ref{fig-7}\textit{c}), and 1.4 (sub-figures \ref{fig-6}\textit{d} and \ref{fig-7}\textit{d}). The dimensionless power terms displayed against the dimensionless time $t^{\ast} = t/t_b$ in the top row of figures \ref{fig-6} and \ref{fig-7} are defined below:
\begin{equation}
{\dot{SE}}^{\ast} =  \frac{1}{E(t=t_0)} \frac{\partial}{\partial t^{\ast}}   \int_{S}^{} \sigma dS ~~~ \text{(surface)}\, ,
\label{eq.SE}
\end{equation}
\begin{equation}
{\dot{KE}}^{\ast} = \frac{1}{E(t=t_0)}  \frac{\partial}{\partial t^{\ast}} \int_{V}^{} \frac{\rho {|\boldsymbol{u}|}^2}{2} dV ~~~ \text{(kinetic)}\, ,
\label{eq.KE}
\end{equation}
\begin{equation}
{\dot{W}}^{\ast} = \frac{1}{E(t=t_0)}  \frac{\partial}{\partial t^{\ast}} \int_{t}^{} \int_{V}^{} \frac{\eta {| \boldsymbol{{\dot{\gamma}}} |}^2 }{2} dV dt ~~~ \text{(dissipated)}\,
\label{eq.W}
\end{equation}
\begin{equation}
{\dot{G}}^{\ast} = \frac{1}{E(t=t_0)}  \frac{\partial}{\partial t^{\ast}} \int_{V}^{} \rho g z dV ~~~ \text{(gravitational)}\, ,
\label{eq.G}
\end{equation}
where $V$ and $S$ respectively represent the volume and the surface of the free-surface part of the liquid (depicted in blue in figure \ref{fig-3}), and $E(t=t_0)$ denotes the total energy of the free-surface part of the liquid at $t = t_0$. The power terms (upper-row sub-figures) are plotted together with the corresponding $d(t)/d_0$ profiles (mid-row sub-figures) and the dimensionless volume-averaged deformation rate terms ${\dot{\gamma}}^{\ast}_{e}$ and ${\dot{\gamma}}^{\ast}_{s}$ (bottom-row sub-figures) defined as  
\begin{equation}
{\dot{\gamma}}^{\ast}_{e} = \frac{1}{V} \int_{V}^{} \frac{1}{2 {|\boldsymbol{\dot{\gamma}}|}^2} \left( {\dot{\gamma}}^{2}_{xx} + {\dot{\gamma}}^{2}_{yy} + {\dot{\gamma}}^{2}_{zz} \right) dV \, ,
\label{eq.gamma-dot-e}
\end{equation}
and
\begin{equation}
{\dot{\gamma}}^{\ast}_{s} = 1 - {\dot{\gamma}}^{\ast}_{e}  \, .
\label{eq.gamma-dot-s}
\end{equation} 
Therefore, ${\dot{\gamma}}^{\ast}_{e}$ (magenta stars) is exclusively linked with extension/compression, while ${\dot{\gamma}}^{\ast}_{s}$ (sky blue crosses) is shear-based.

\begin{figure*}[htp]
\centering    
\includegraphics[width=1\linewidth]{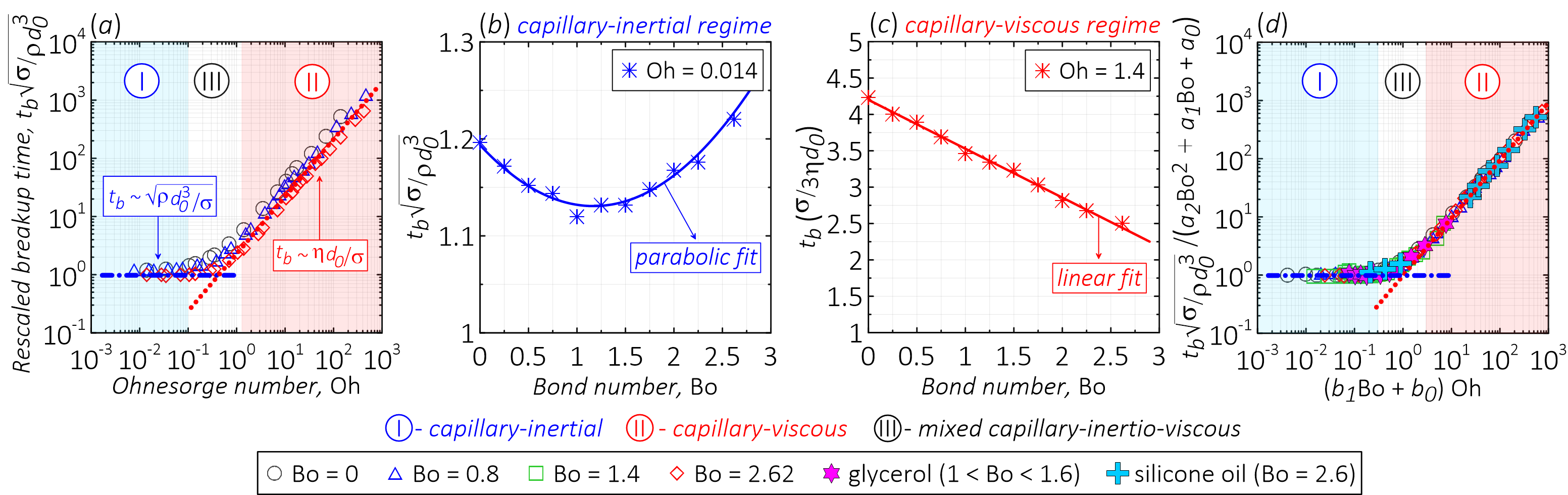}
\vspace{-0.5cm}    
\caption{{\color{darkgray}{(\textit{a}) Numerical results showing the rescaled breakup time $t_b \sqrt{\sigma/(\rho d_0^3)}$ (e.g., the breakup time made dimensionless by the characteristic capillary-inertial time given by equation \ref{eq.tb-ci}) as a function of Ohnesorge number for numerical results (open symbols) at three Bond numbers: $\mathrm{Bo} = 0$ (grey circles), $\mathrm{Bo} = 0.8$ (blue triangles), and $\mathrm{Bo} = 2.62$ (red diamonds). The data collapse across a master curve divided into three regions highlighting different capillary-thinning regimes: capillary-inertial (I; in blue), for which $t_b \sim \sqrt{\rho d_0^3/\sigma}$; capillary-viscous (II; in red), where $t_b \sim \eta d_0/\sigma$; and a mixed regime (III; in white), for which capillary, inertial and viscous effects all play an essential role. (\textit{b}) Numerical results underlining Bond number effects on the breakup time (made dimensionless by the characteristic capillary-inertial time $t_{ci} = \sqrt{\rho d_0^3/\sigma}$) in the capillary-inertial regime ($\mathrm{Oh} = 0.014$). The data is fitted by a quadratic polynomial $a_2 {\mathrm{Bo}}^2 + a_1 {\mathrm{Bo}} + a_0$, with $a_2 = 0.0458$, $a_1 = -0.1099$, and $a_0 = 1.1958$ (blue line). (\textit{c}) Numerical results highlighting Bond number effects on the breakup time (made dimensionless by the characteristic capillary-viscous time $t_{cv} = 3 \eta d_0/\sigma$) in the capillary-viscous regime ($\mathrm{Oh} = 1.4$). The numerical points follow a linear decay described as $b_1 {\mathrm{Bo}} + b_0$, with $b_1 = -0.6667$, $b_0 = 4.1964$ (red line). (\textit{d}) The rescaled time, divided by the parabolic fit, is plotted as a function of the Ohnesorge number, multiplied by the linear fit, for numerical and experimental results (open and solid symbols, respectively).}}}
\vspace{-0.2cm}  
\label{fig-8}
\end{figure*}

For all cases shown in figure \ref{fig-6} ($\mathrm{Bo} = 0$), the filament breakup is exclusively induced by the capillary pressure in a gravity-free context (${\dot{G}}^{\ast} = 0$). Hence, the surface energy (grey circles) is converted into kinetic energy (blue triangles) and dissipated (red diamonds) by viscous effects. At $\mathrm{Oh} = 0.014$ (sub-figure \ref{fig-6}\textit{a}), ${\dot{SE}}^{\ast}$ is primarily converted into ${\dot{KE}}^{\ast}$, indicating that the filament thinning process is dominated by the balance between capillary and inertial stresses (note that ${\dot{SE}}^{\ast}$ and ${\dot{KE}}^{\ast}$ are strongly correlated sub-figure \ref{fig-6}\textit{a}, while ${\dot{W}}^{\ast} \approx 0$ and ${\dot{G}}^{\ast} = 0$). This is a capillary-inertial regime, in which the $d(t)/d_0$ curve (black asterisks) exhibits a highly non-linear form proportional to $(1-t/t_b)^{2/3}$, as pointed out by the blue dotted line. As $\mathrm{Oh}$ increases at a fixed $\mathrm{Bo}$, dissipation becomes more pronounced (sub-figures \ref{fig-6}\textit{b}-\textit{c}) eventually dominating the thinning process (sub-figure \ref{fig-6}\textit{d}), while $d(t)/d_0$ becomes mainly linear, e.g., $d(t)/d_0 \propto 1-t/t_b$, as indicated by the red dashed line. This is a capillary-viscous regime. Lastly, although the flow cases at $\mathrm{Bo} = 0$ present essentially an extensional character, they become exclusively extension-dominated only when the capillary-viscous regime is reached (note that ${\dot{\gamma}}^{\ast}_{e} > 0.5$ along the whole filament thinning process only in sub-figure \ref{fig-6}\textit{d}).   

By comparing figures \ref{fig-6} and \ref{fig-7}, we clearly note that gravity effects are not marginal at $\mathrm{Bo} = 2.62$. However, they are not high enough to make the thinning process primarily gravity-driven either. As a result, we still observe a strong correlation between ${\dot{SE}}^{\ast}$ and ${\dot{KE}}^{\ast}$ at $\mathrm{Oh} = 0.014$ (capillary-inertial regime; sub-figure \ref{fig-7}\textit{a}), and ${\dot{SE}}^{\ast}$ and ${\dot{W}}^{\ast}$ at $\mathrm{Oh} = 1.4$ (capillary-viscous regime; sub-figure \ref{fig-7}\textit{d}). Once again, the dimensionless minimum diameter curves appear as being proportional to $(1-t/t_b)^m$, with $m = 2/3$ at $\mathrm{Oh} = 0.014$, and $m = 1$ at $\mathrm{Oh} = 1.4$, which is in line with equation \ref{eq:intro-1}. Nevertheless, as highlighted by the blue and red boxes in figures \ref{fig-6} and \ref{fig-7}, the fitted portion of $d(t)/d_0$ decreases as $\mathrm{Bo}$ becomes more pronounced. Additionally, the slope of the fits varies with $\mathrm{Bo}$ as well, which makes the use of equation \ref{eq:intro-1} for rheological purposes difficult. Finally, although the gravity strengthening favours the increase of ${\dot{\gamma}}^{\ast}_{s}$ (which is particularly related to the liquid flow within the lower droplet), it is worth noting that, in the capillary-viscous regime, the thinning process remains extension-dominated from $0 \leq t/t_b \leq 1$ for the whole range of $\mathrm{Bo}$ explored here ($\mathrm{Bo} \leq 3$).                                         

It is crucial to emphasise that, although just a part of the diameter curves in sub-figures \ref{fig-6}(\textit{a}) and \ref{fig-7}(\textit{a}), and sub-figures \ref{fig-6}(\textit{d}) and  \ref{fig-7}(\textit{d}) can be fitted by equation \ref{eq:intro-1}, the whole thinning process is dominated by a balance between capillary and inertial stresses, and capillary and viscous stresses, respectively. Such observations encourage us to explore the link connecting the breakup time and the physical mechanisms driving the capillary-thinning process. Hence, based on the energy transfer analyses conducted above, we propose scaling laws for $t_b$ by considering the capillary-inertial regime (sub-figures \ref{fig-6}\textit{a} and \ref{fig-7}\textit{a}), and the capillary-viscous regime (sub-figures \ref{fig-6}\textit{d} and \ref{fig-7}\textit{d}).

In the capillary-inertial thinning regime, the filament surface energy ($\sim \sigma d_0^2$) is
primarily converted into kinetic energy ($\sim \rho u_c^2 d_0^3$, where $u_c$ is the characteristic filament thinning velocity approximated as $u_c \sim d_0/t_b$), leading to 
\begin{equation}
t_{ci} = \sqrt{\frac{\rho d_0^3}{\sigma}} ~~~ \text{(capillary-inertial time)} \, ,
\label{eq.tb-ci}
\end{equation}
a well-known characteristic capillary-inertial time extensively discussed in previous works \citep{Peregrine_1990, Eggers_1993, Eggers_2005, Javadi_2013, Deblais_2018, Dinic_2019, Bazazi_2025}. On the other hand, in the capillary-viscous regime, the filament surface energy is mainly dissipated by viscous effects through an extension-dominated process ($\sim \eta {\dot{\gamma}}_c d_0^3$, in which ${\dot{\gamma}}_c$ is the characteristic filament thinning velocity estimated as ${\dot{\gamma}}_c \sim 1/t_b$). Hence, we find 
\begin{equation}
t_{cv} = \frac{3 \eta d_0}{\sigma} ~~~ \text{(capillary-viscous time)} \, ,
\label{eq.tb-cv}
\end{equation} 
an equally well-known characteristic capillary-viscous breakup time vastly discussed in the literature \citep[the prefactor 3 emerges from the uniaxial extensional nature of the flow underlined by sub-figures \ref{fig-6}\textit{d} and \ref{fig-7}\textit{d}; see also][]{Papageorgiou_1995, McKinley_2000, Anna_2001}. By equating the above characteristic times (equations \ref{eq.tb-ci} and \ref{eq.tb-cv}), we find $\mathrm{Oh} \sim 1$. In other words, $\mathrm{Oh}$ tends to 1 when transitioning from the capillary-inertial thinning regime to the capillary-viscous one. 

The validity of the above theoretical arguments is corroborated by sub-figure \ref{fig-8}(\textit{a}) in which the rescaled breakup time $t_b \sqrt{\sigma/\rho d_0^3}$ (e.g., the breakup time made dimensionless by the characteristic capillary-inertial breakup time given by equation \ref{eq.tb-ci}) is plotted as a function of Ohnesorge number for numerical results (open symbols) at three Bond numbers: $\mathrm{Bo} = 0$ (grey circles), $\mathrm{Bo} = 0.8$ (blue triangles), and $\mathrm{Bo} = 2.62$ (red diamonds). The results collapse across a master curve divided into three regions highlighting different capillary-thinning regimes: capillary-inertial (I; in blue), for which $t_b \sim \sqrt{\rho d_0^3/\sigma}$; capillary-viscous (II; in red), where $t_b \sim \eta d_0/\sigma$; and a mixed regime (III; in white), for which capillary, inertial and viscous effects all play an essential role. As underlined by our theoretical analyses, a mixed regime (transitioning zone) emerges when $ 0.1 < \mathrm{Oh} < 1.1$. 

Figure \ref{fig-8}(\textit{a}) also reveals slight gravitational effects on $t_b$, which can be perceived by comparing its different symbols. The increase in $\mathrm{Bo}$ speeds up the breakup process, a direct consequence of the increased gravity stresses applied to the liquid filament. The gravitational effects on $t_b$ are quantified in sub-figures \ref{fig-8}(\textit{b}) and \ref{fig-8}(\textit{c}) for both the capillary-inertial and the capillary-viscous regimes by plotting respectively $t_b \sqrt{\sigma/\rho d_0^3}$ versus $\mathrm{Bo}$ at $\mathrm{Oh} = 0.014$ (sub-figure \ref{fig-8}\textit{b}), and $t_b \sigma/3 \eta d_0$ versus $\mathrm{Bo}$ at $\mathrm{Oh} = 1.4$ (sub-figure \ref{fig-8}\textit{c}). Note that $t_b \sqrt{\sigma/\rho d_0^3}$ is the breakup time made dimensionless by the characteristic capillary-inertial breakup time, while $t_b \sigma/3 \eta d_0$ denotes the breakup time made dimensionless by the characteristic capillary-viscous breakup time. These dimensionless times can be respectively fitted by a quadratic polynomial ($a_2 {\mathrm{Bo}}^2 + a_1 {\mathrm{Bo}} + a_0$, with $a_2 = 0.0458$, $a_1 = -0.1099$, and $a_0 = 1.1958$; blue line in sub-figure \ref{fig-8}\textit{b}) and a linear one ($b_1 {\mathrm{Bo}} + b_0$, with $b_1 = -0.6667$, $b_0 = 4.1964$; red line in sub-figure \ref{fig-8}\textit{c}). The parabolic fit is valid for flow cases at $\mathrm{Oh} \le 0.1$ and $\mathrm{Bo} \le 3$ (blue region in \ref{fig-8}\textit{a}), while the linear fit is valid for $\mathrm{Oh} \ge 1.1$ and $\mathrm{Bo} \le 3$ (red region in \ref{fig-8}\textit{a}). Hence, the mentioned fitting curves can be used to make the dimensionless quantities employed in \ref{fig-8}(\textit{a}) able to capture gravitational effects for $\mathrm{Bo} \leq 3$. This is highlighted in figure \ref{fig-8}(\textit{d}), where the rescaled time divided by the parabolic fit is plotted as a function of the Ohnesorge number multiplied by the linear fit for numerical and experimental results (open and solid symbols, respectively). The data collapse perfectly onto a single path, which can ultimately be used to estimate either the viscosity or the surface tension of Newtonian fluids based on their $t_b$. In other words, the master curve illustrated in sub-figure \ref{fig-8}\textit{d}) can be used either as a viscosimeter or a tensiometer for Newtonian fluids (and $\mathrm{Bo} \leq 3$).                  

\begin{figure*}[htp]
\centering    
\includegraphics[width=1\linewidth]{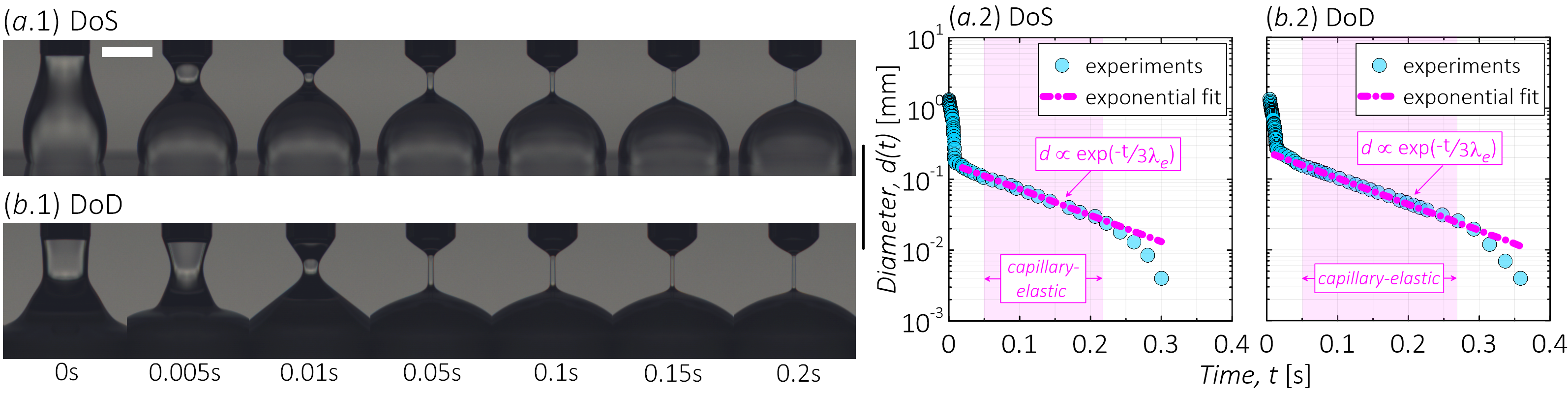}
\vspace{-0.7cm}    
\caption{{\color{darkgray}{DoS \textit{versus} DoD. (\textit{a}.1 and \textit{b}.1) Experimental snapshots illustrating the capillary thinning process of a $5\times10^6$g/mol-molecular-weight polyethylene oxide solution (PEO; provided by Sigma-Aldrich) of 100 parts per million (ppm) by weight in deionised water using DoS (\textit{a}.1) and DoD (\textit{b}.1). White scale bar = 1.37mm. (\textit{a}.2 and \textit{b}.2) The snapshots are converted into curves showing the filaments' minimum diameter as a function of time for DoS (\textit{a}.2) and DoD (\textit{b}.2). As underlined by the exponential fits in the pink boxes (dashed-pink lines), DoS and DoD lead to the same $\lambda_e =$ 39ms. The experiments are performed at $\mathrm{Bo} = 1.3$. The error bars are comparable in size to the symbols. Note that, to favour straightforward comparisons between DoS and DoD, only the thinning process ($d(t) \leq d_0$) is illustrated for the DoD case.}}}
\vspace{-0.2cm}  
\label{fig-9}
\end{figure*}

Finally, figure \ref{fig-9} indicates that, similar to the Dripping-onto-Substrate capillary breakup \citep[DoS; e.g., a nozzle-generated pendant drop spreads upon a solid creating a fluid filament that subsequently breaks apart under capillary stresses;][]{Dinic_2017, Zinelis_2024}, the DoD method can be used to estimate the extensional relaxation time $\lambda_{e}$ of polymer solutions. More specifically, figure \ref{fig-9} compares through snapshots (left side) and minimum diameter curves (right side) the breakup dynamics and the extensional relaxation time of a $5\times10^6$g/mol-molecular-weight polyethylene oxide solution \citep[PEO; provided by Sigma-Aldrich; see also][]{Pereira_2013} of 100 parts per million (ppm) by weight in deionized water using DoS (sub-figures \ref{fig-9}\textit{a}.1-\ref{fig-9}\textit{a}.2) and DoD (sub-figures \ref{fig-9}\textit{b}.1-\ref{fig-9}\textit{b}.2). The experiments are carried out at $\mathrm{Bo} = 1.3$. Extensional relaxation time values are obtained by fitting the diameter curves with equation \ref{eq:intro-2} (emerging from the capillary-elastic thinning regime). As underlined by the exponential fits in the pink boxes (dashed-pink lines), DoS and DoD lead to the same $\lambda_e =$ 39ms. 

\section{Concluding Remarks}

We have presented here a mixed experimental, numerical, and theoretical study of the physical mechanisms governing the stretching and subsequent breakup of a Newtonian fluid filament generated after the coalescence of an upper millimetric droplet with a lower droplet cap. Hence, in such a wetting-independent configuration, the capillary thinning process naturally occurs when the upper drop gently touches the lower one (no fast pre-stretching required). Moreover, the absence of axial displacement during the liquid filament formation mitigates the development of secondary inertial stresses throughout the capillary-thinning stage. Our mixed approach combined filament breakup experiments recorded by a high-speed camera with three-dimensional (3D) numerical simulations based on a variational multiscale framework for multiphase fluid flows. The results were analysed in light of stretching dynamics, energy budgets, and scaling laws. To extend the classical approach based on the time evolution of the filament diameter, we also examined the relationship between the breakup time and the physical mechanisms driving the thinning process. 

Three flow regimes are highlighted: the capillary-inertial, resulting from a competition between capillary and inertial stresses, for which the breakup time $t_b$ scales with $\sqrt{\rho d_0^3/\sigma}$;  the capillary-viscous regime, emerging from a balance between capillary and viscous stresses, for which $t_b \sim \eta d_0/\sigma$; and the mixed capillary-inertial-viscous for which capillary, inertial and viscous are all important. These regimes are secondarily affected by gravitational stresses even for $\mathrm{Bo} < 3$. Such effects on $t_b$ are translated by a first-order and a second-order polynomial. As a result, the breakup time is described by a master curve that depends on both the Ohnesorge and Bond numbers. The master curve can ultimately be used either as a viscosimeter or a tensiometer for Newtonian fluids (for $\mathrm{Bo} \leq 3$). 

Finally, we have briefly shown that DoD can equally give access to the extensional relaxation time of polymer solutions. It would thus be interesting to extend these analyses in future works by considering a wider selection of polymers. Another upstanding possibility would be using DoD in a three-phase-flow configuration to study, for instance, mass-transfer and/or ionic-gelation effects on the capillary-thinning process, i.e., the upper polymer-based droplet undergoes ion-induced gelation following its coalescence with the lower droplet containing ions \citep[the polymer molecules crosslink as the ions diffuse within the fluid filament;][]{Godefroid_2025}.

\textit{\textbf{Acknowledgements:}} 
This research is co-funded by The Transition Institute 1.5 of Mines Paris - PSL. The authors would also like to acknowledge the support of the PSL Research University under the program `Investissements d'Avenir' launched by the French Government and implemented by the French National Research Agency (ANR) with the reference ANR-10-IDEX-0001-02 PSL, and the ANR for supporting the INNpact project (ANR-21-CE06-0036) under the `Jeunes chercheuses et jeunes chercheurs' program.    


\bibliography{Newtonian_DoD-ref}
\end{document}